\documentclass[aps,twocolumn,pra,floatfix,superscriptaddress]{revtex4-1}
\usepackage{times}
\usepackage{amsmath,amssymb}
\usepackage{graphicx}
\usepackage{xcolor}
\graphicspath{{graphics/}}
\usepackage{braket}
\usepackage{float}
\usepackage{tikz}
\usepackage[pdftex, colorlinks, allcolors=blue]{hyperref}

\newcommand{\zeropi}{$0$-$\pi$\,}
\newcommand{\dket}[1]{\left|#1\right\rangle\! \rangle}

\begin{document}

\title{Universal gates for protected superconducting qubits using optimal control}

\author{Mohamed Abdelhafez}
\affiliation{The James Franck Institute and Department of Physics, University of Chicago, Chicago, Illinois 60637, USA}

\author{Brian Baker}
\affiliation{Department of Physics and Astronomy, Northwestern University, Evanston, Illinois 60208, USA}

\author{Andr\'as Gyenis}
\affiliation{Department of Electrical Engineering, Princeton University, Princeton, New Jersey 08544, USA}

\author{Pranav Mundada}
\affiliation{Department of Electrical Engineering, Princeton University, Princeton, New Jersey 08544, USA}

\author{Andrew A.\ Houck}
\affiliation{Department of Electrical Engineering, Princeton University, Princeton, New Jersey 08544, USA}

\author{David Schuster}
\affiliation{The James Franck Institute and Department of Physics, University of Chicago, Chicago, Illinois 60637, USA}

\author{Jens Koch}
\affiliation{Department of Physics and Astronomy, Northwestern University, Evanston, Illinois 60208, USA}

\begin{abstract}
We employ quantum optimal control theory to realize quantum gates for two protected superconducting circuits: the heavy-fluxonium qubit and the \zeropi qubit. Utilizing automatic differentiation facilitates the simultaneous inclusion of multiple optimization targets, allowing one to obtain high-fidelity gates with realistic pulse shapes. For both qubits, disjoint support of low-lying wave functions prevents direct population transfer between the computational-basis states. Instead, optimal control favors dynamics involving higher-lying levels, effectively lifting the protection for a fraction of the gate duration. For the \zeropi qubit, offset-charge dependence of matrix elements among higher levels poses an additional challenge for gate protocols. To mitigate this issue, we randomize the offset charge during the optimization process, steering the system towards pulse shapes insensitive to charge variations. Closed-system fidelities obtained are 99\% or higher, and show slight reductions in open-system simulations.
\end{abstract}
\maketitle

\section{Introduction\label{Introduction}}
The ability to perform fast, high-fidelity gate operations on qubits is critical for quantum information processing. A host of research over the last decades has pursued optimal strategies to realize qubit gates \cite{Khaneja_2005,Glaser2015Training,deFouquieres2011Second,Sklarz2002Loading,Eitan2011Optimal,Gollub2008Monotonic,Nigmatullin2009Implementation,Reich2012Monotonically,Palao2003Optimal,Maday2003New,Borzi2008Formulation,Ditz2008Cascadic}. Realizing universal sets of quantum gates has been achieved through a variety of techniques for different qubit implementations \cite{Sleator1995,zhu2002implementation,zu2014experimental,schmidt2003realize,monz2009realization}. 
One standard example of a minimal universal set of gates consists of the single-qubit Hadamard gate H, the single-qubit T gate and one two-qubit entangling gate such as the controlled-Z gate. Realizing these building blocks with high fidelities is crucial in order to meet the gate-fidelity threshold required for error correction codes \cite{Chow2012,nielsen00}.

With superconducting qubits, optimized gates with fidelities exceeding 99$\%$ have been proposed \cite{Chow2009,Chow2012,chow2010optimized,Leung_2017,Open-Mohamed}, for example for the transmon qubit \cite{Koch2007,Schreier2008}. 
Driving transitions in the computational subspace of the transmon qubit is facilitated by direct matrix elements between the states $\ket{0}$ and $\ket{1}$ whose wavefunctions reside in the same cosine-potential well. Recently, a new generation of superconducting qubits has been introduced which feature disjoint support: low-energy wavefunctions are localized in different potential wells so that matrix elements of local operators are exponentially suppressed \cite{Kitaev2006,Brooks2013,Earnest2018,Vool2018}. Therefore, these qubits are intrinsically protected from spontaneous transitions between the computational-basis states. Two of the most promising protected superconducting qubits are the heavy-fluxonium qubit \cite{Manucharyan2009,Earnest2018,Nguyen2018} and the \zeropi circuit \cite{Brooks2013,Dempster2014,Groszkowski_2018,DiPaolo2019}. Performing high-fidelity gates on these qubits is challenging precisely because of the lack of direct transition matrix elements. For the heavy-fluxonium qubit, these forbidden transitions have been successfully accessed by stimulated Raman processes. \cite{Earnest2018,Vool2018}. For the \zeropi circuit, a recent study has proposed DC-voltage signals for realizing either an X gate or Hadamard gate \cite{DiPaolo2019}. Here, we argue that optimal-control theory is a promising route to explore the options for high-fidelity gates in protected qubits such as heavy fluxonium and \zeropi.

Optimal-control theory, applied to quantum systems, achieves a set of optimization targets, the primary target usually consisting of a maximized gate or state-transfer fidelity. Additional constraints associated with specific experimental systems may be added, and include smoothing of control pulses and limiting their amplitudes \cite{Skinner2004Reducing,Kobzar2004Exploring,Kobzar2008Exploring}, as well as accounting for the limited time resolution of arbitrary waveform generators \cite{Motzoi2011Optimal}. There are many different implementations of optimal-control algorithms. Examples of such algorithms include implementations for closed \cite{Khaneja_2005,deFouquieres2011Second,Eitan2011Optimal,Gollub2008Monotonic,Leung_2017} and open \cite{Khaneja_2005,boutin2017resonator,Open-Mohamed} quantum systems, most of them are gradient based. Some of these algorithms are available as open-source packages \cite{Johansson2013QuTiP,Machnes2011Comparing,Hogben2011Spinach}, and we here utilize the automatic-differentiation \cite{Baydin_Pearlmutter_Radul_Siskind_2015} based quantum optimizer we previously introduced in Ref.\ \onlinecite{Leung_2017}. Automatic differentiation allows for the flexibility of adding optimization targets without calculating their analytical gradients. We utilize and further develop this optimal-control implementation to obtain a universal set of gates for the protected heavy-fluxonium and \zeropi qubits. For the latter, we find that optimal-control pulses strikingly succeed in overcoming the obstacle of offset-charge dependent matrix element.

\section{Optimal Control Theory\label{OCT}}
Quantum optimal control helps steer the time evolution of quantum systems to realize a desired state transfer, unitary operation, or readout protocol \cite{Skinner2004Reducing,Kobzar2004Exploring,Chen2015Neartimeoptimal}.  This is accomplished by optimizing a set of external control pulses $\{u_1(t),\ldots,u_M(t)\}$ which couple to the quantum system via control operators $\{\mathcal{H}_1,\ldots,\mathcal{H}_M\}$ and, thus, change the system dynamics. The resulting time-dependent Hamiltonian has the general form
\begin{equation}
H(t) = \mathcal{H}_0 + \sum_{k=1}^M u_{k}(t)\mathcal{H}_{k},
\label{eq:Hamiltonian}
\end{equation}
where $\mathcal{H}_0$ is the intrinsic system Hamiltonian, also known as drift Hamiltonian. 
The task of optimization is to determine a set of control pulses which minimize a cost functional $C[\{u_k(t)\}]$. This functional encodes the infidelity of the target process, and may include additional optimization constraints crucial for achieving realistic pulses. 

We briefly review the pertinent contributions to the cost functional employed in our work. 
In the case of a target unitary operation $U_t$, acting on a closed system, the primary cost to be minimized is the gate infidelity 
 \begin{equation}
     C_1 = 1- F_c = 1- \frac{1}{n^2}|\text{Tr}(U_t^{\dagger} U_f)|^2.
 \end{equation}
 Here, $U_f$ is the unitary realized by a given set of control pulses, and $n$ denotes the dimension of the Hilbert space. Secondary optimization targets are utilized to smooth control pulses and limit their signal power so to enable their implementation in the laboratory setting. In addition, cost penalties for occupation of certain higher-lying states help avoid leakage and ensure the validity of the inevitable Hilbert-space truncation. The individual contributions to the cost functional are summarized in Table \ref{table:cost_fns}.
 \begin{table}[htbp]
\caption{Relevant contributions to the cost functional. Indices $k$, $j$ label the $k$-th control pulse and the $j$-th discretized time step.} 
\centering 
\begin{ruledtabular}
\begin{tabular}{c p{0.5\linewidth} } 
 Cost functional contribution &  \centering Explanation  \tabularnewline 
\hline 
$C_1 = 1- \frac{1}{n^2}|\text{Tr}(U_t^{\dagger} U_f)|^2$ &  Infidelity of realized unitary $U_f$ relative to target gate $U_t$ \\
$C_2 = \sum_j |\langle\psi_f|\psi_{j}\rangle|^2$ &  Occupation of forbidden state $\psi_f$\\ 
$C_3 = \sum_{k, j} | u_{kj} - u_{k j-1} |^2$ &  First derivatives of the control parameters\\
$C_4 = \sum_{k,j} |u_{kj}|^2$ & Control pulse power 
\end{tabular}
\label{table:cost_fns} 
\end{ruledtabular}
\end{table}
 
 A common technique for cost minimization consists of Gradient Ascent Pulse Engineering (GRAPE) \cite{Khaneja_2005} based on explicit analytical expressions for the gradients of $C$ with respect to each of the control pulses $u_k(t)$. Here, we instead utilize an automatic-differentiation optimizer \cite{Leung_2017} built on TensorFlow \cite{Abadi_Others_2016}. This avoids the need for hard-coded gradients of each new contribution to the cost functional.
 
To assess gate fidelities in the presence of dissipation and dephasing, we employ a Lindblad master equation description \cite{BRE02} of the quantum system weakly interacting with its environment,
\begin{equation}
\frac{d\rho}{dt} = - i [H,\rho] + \sum_l \gamma_l \big[ c_l \rho c_l^{\dagger} -\tfrac{1}{2}\{ c_l^{\dagger} c_l, \rho\} \big] .
\end{equation}
Here, $\rho$ is the reduced density matrix of the system, and $\{c_l\}$ a set of jump operators capturing relaxation and dephasing processes with associated rates $\{ \gamma_l \}$. The metric we use for open-system gate fidelity is given by \begin{equation}
    F_o = \frac{1}{n^2} \text{Tr} (L_t^{\dagger} L_f),
\end{equation}
where $L_t = U_t  \otimes U_t^* $ is the target superoperator and $L_f$ is the final superoperator defined by  $L_f \rho(0) = \rho(t)$, i.e. it propagates a vectorized version of the system density matrix (see Appendix \ref{fidelitycalculation} for more details).

\section{Optimized universal-gate set for the heavy fluxonium qubit\label{fluxonium}}
\subsection{Single-qubit gates}\label{single_qubit_gates}
The fluxonium qubit \cite{Manucharyan2009} is a promising superconducting circuit that may, in its most recent variants as ``heavy fluxonium''  \cite{Manucharyan2009,Earnest2018}, outperform the widely used transmon qubit \cite{Koch2007,Schreier2008}. In contrast to the transmon, heavy fluxonium combines strong Josephson non-linearity with  $T_1$ protection due to disjoint support of its lowest-lying localized  wave functions. Heavy fluxonium devices utilize a decreased capacitive energy $E_C$, which emphasizes the localization of states \cite{Manucharyan2009,Earnest2018}. Moreover, fluxonium eigenenergies are intrinsically insensitive to slow offset charge variations \cite{Koch2009}. The protection granted by disjoint state support, however, also complicates the realization of universal gate operations by means of external microwave pulses: matrix elements for direct transitions between disjoint-support states remain exponentially suppressed. In this section, we show that optimal control algorithms can nevertheless yield efficient protocols for a universal gate set. Such protocols necessitate involvement of higher qubit levels, and we carefully evaluate fidelity limitations arising from temporary occupation of these states.
 
 Experimentally, gates for heavy fluxonium have been realized by driving Raman transitions \cite{Vool2018,Earnest2018}, which utilize intermediary higher-energy states to assist indirect transitions between the protected states. We will demonstrate a similar approach, exploiting the availability of intermediary state transitions using optimal control theory. The optimal-control formalism offers greater flexibility in terms of pulse shape, and yields fast, high-fidelity single-qubit gates with gate times below 100\,ns and fidelities exceeding 99.9\%. We obtain optimized pulse shapes for X, H, and T gates, thereby establishing a blueprint for  realizing arbitrary single-qubit gate operations.
 
 \begin{figure}
  \includegraphics[width=0.75\linewidth]{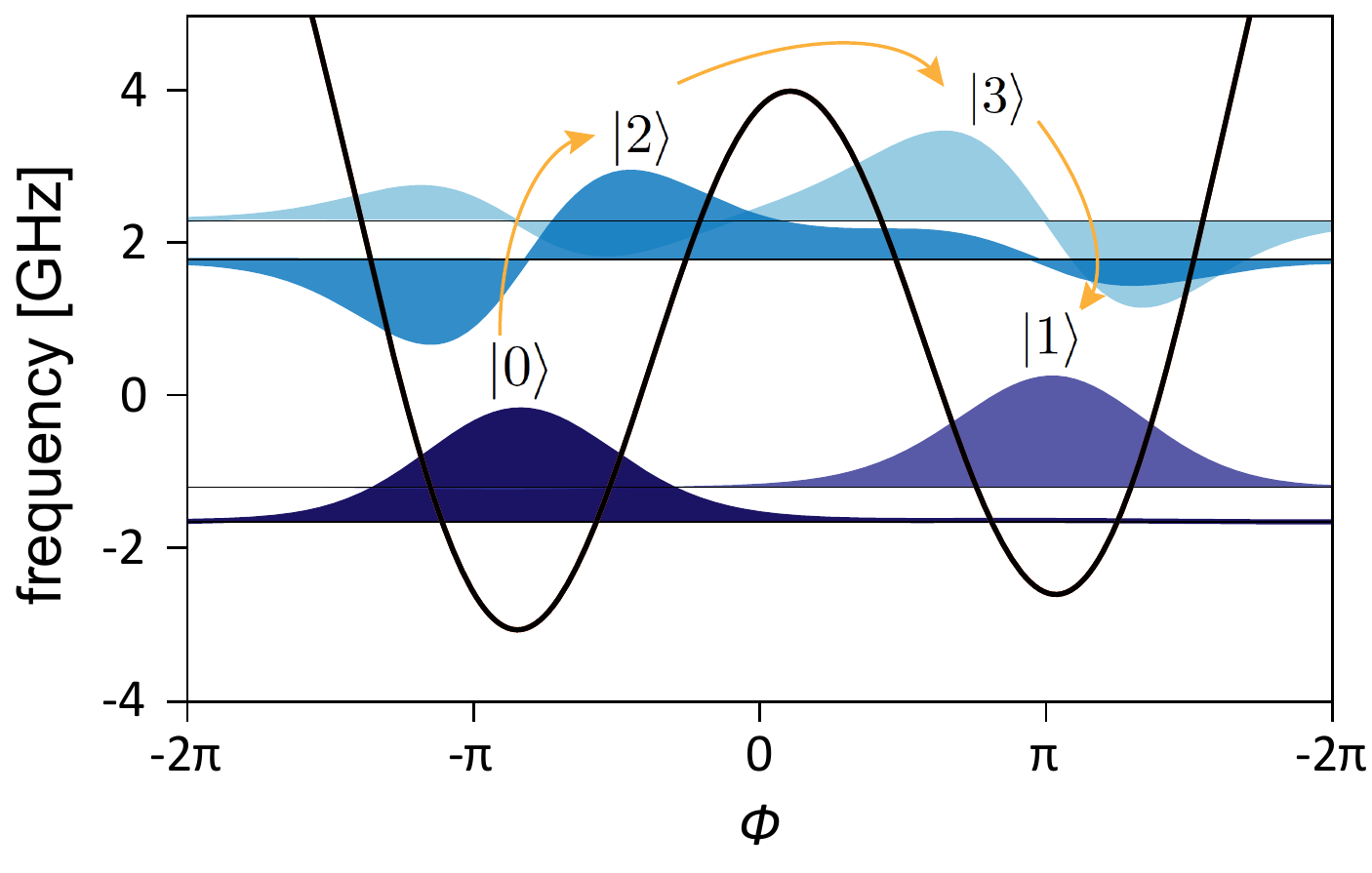}\\[-4mm]
  \caption{First four fluxonium wave functions, slightly away from the flux sweet spot ($\Phi_{\text{ext}}=0.45\Phi_0$). The lowest-lying states $\ket{0}$ and $\ket{1}$ are localized and have practically disjoint support. The auxiliary states $\ket{2}$ and $\ket{3}$ delocalize over both potential wells and serve as intermediate states for quantum gates. Gates involving population transfer between $\ket{0}$ and $\ket{1}$ such as X or H gates utilize the delocalized states for transfer across the potential barrier.
\label{fig:fluxonium_potential}}
\end{figure}
 
 As typical in circuit QED \cite{Wallraff2004,Blais2004}, each gate is realized by a microwave pulse applied to a transmission-line resonator which, in turn, is coupled to the qubit. The corresponding Hamiltonian for this driven, generalized Jaynes-Cummings model is
 \begin{align}\nonumber 
     H_{JC} &= \sum_{l}\epsilon_l\ket{l}\!\bra{l} + \omega_r a^\dag \!a + \sum_{l,l'}g_{ll'}\ket{l}\!\bra{l'}(a^\dag + a)\\
     &\quad + v(t)(a^\dag + a),
     \label{res_flux_Ham}
 \end{align}
where $\epsilon_l$, $\ket{l}$ are the fluxonium eigenenergies eigenstates labeled by index $l$, and $\omega_r$ is the resonator frequency. The relative coupling strengths are given by $g_{ll'} = g\langle l | n_\phi | l'\rangle$, where $n_\phi$ is the fluxonium charge operator. Fluxonium eigenenergies and eigenstates are governed by the Hamiltonian \cite{Manucharyan2009,Koch2009}
\begin{equation}
    H_f = 4E_C n_\phi^2 + \frac{1}{2}E_L\phi^2 - E_J\text{cos}(\phi + 2\pi\Phi_{\text{ext}}/\Phi_0),
\end{equation}
in which $E_C$, $E_L$, and $E_J$ denote the capacitive, inductive, and Josephson energies, respectively. $\Phi_{\text{ext}}$ is the external magnetic flux threading the loop formed by the junction and inductor.
For the heavy-fluxonium qubit, we choose realistic device parameters $E_C/h = 0.5\,$GHz , $E_L/h = 0.25\,$GHz, and $E_J/h = 4.0\,$GHz, and a flux working point slightly away from half-integer flux, $\Phi_{\text{ext}} = 0.45\Phi_0$. This places the system in the protected regime of nearly degenerate states  $\ket{0}$ and $\ket{1}$ with disjoint support, see Fig.~\ref{fig:fluxonium_potential}. (Operating the qubit away from the half-integer flux sweet spot increases sensitivity to dephasing from $1/f$ flux noise, which we monitor closely in our analysis.)

Throughout this work, we focus on dispersive control of the qubit, in which the drive tone $v(t)$ steers dynamics within the qubit subsystem, but leaves the resonator state essentially unchanged. This allows us to exclude the resonator subspace from explicit simulation within the optimal-control algorithm. (We verify in a separate simulation that the resonator state is unaffected by the drive tone, i.e.\ the average photon number obeys $\langle a^\dag\!a \rangle \ll 1$ throughout the evolution.) In the resulting driven-fluxonium Hamiltonian
\begin{equation}\label{total_Ham_single_qubit}
    H(t) = H_f + V(t),     
\end{equation}
we properly account for the fact that the drive on the qubit is filtered through the resonator. The dispersive coupling between qubit and resonator produces an effective drive on the qubit of the form
\begin{equation}\label{filtered_drive}
    V(t) = 2g\omega_r v(t) \sum_{l,l'}\frac{\bra{l}n_\phi\ket{l'}}{(\epsilon_{l} - \epsilon_{l'})^2-\omega_r^2}\ket{l}\!\bra{l'},
\end{equation}
see Appendix \ref{dispersive} for details. For our simulation, we consider a coupling strength and resonator frequency of $g/2\pi = 300$ MHz and $\omega_r/2\pi = 7.5$ GHz, respectively. 

\begin{figure*}
  \includegraphics[width=\linewidth]{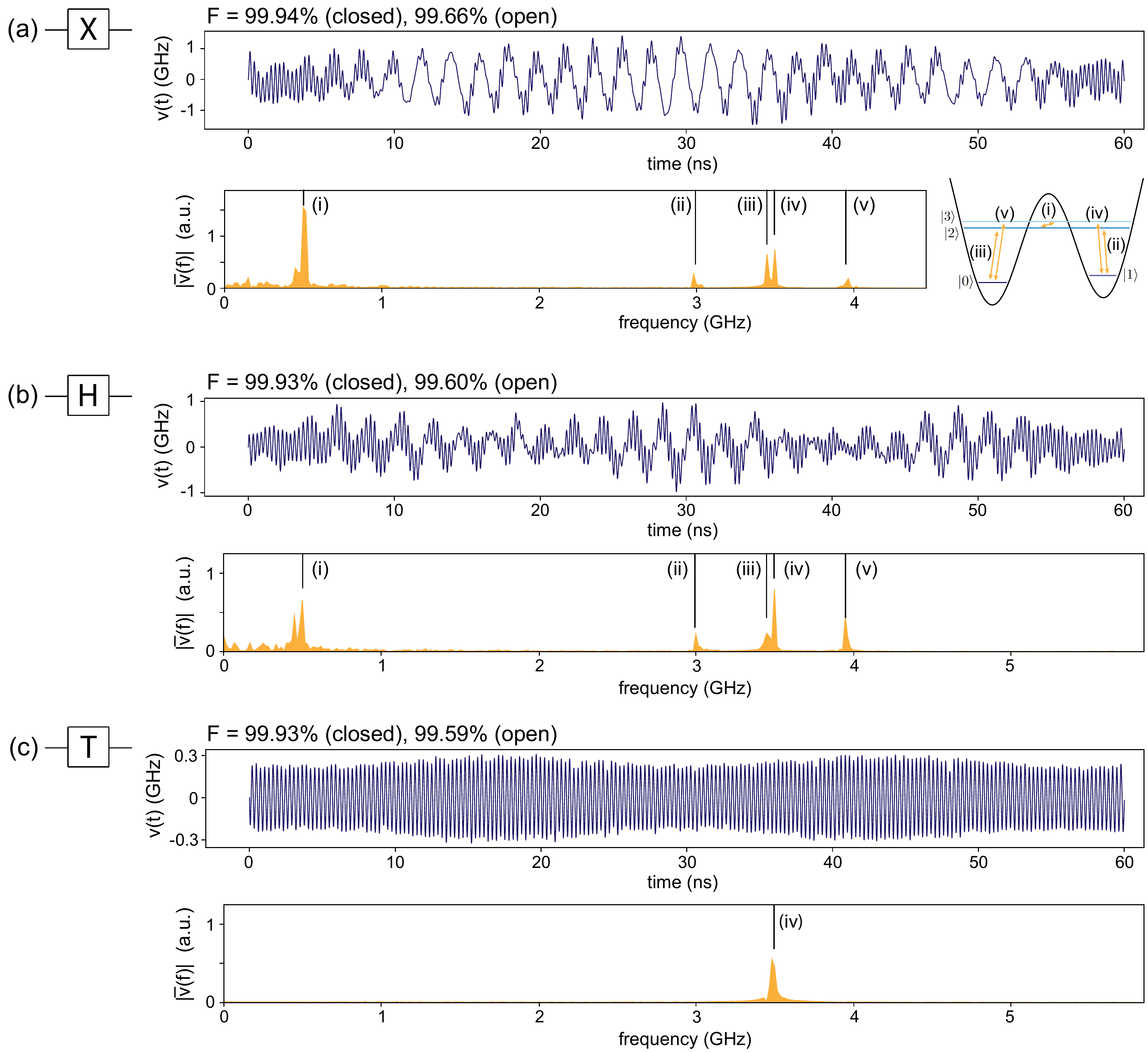}
  \caption{High-fidelity single-qubit gates for heavy fluxonium. (a) Optimized pulse shape $v(t)$ and its discrete Fourier transform $\bar{v}(f)$ for the Pauli-X gate, achieving a gate fidelity of 99.94\%. The Fourier transform exhibits distinct peaks that align with the transition frequencies among the involved levels (see inset). (b) Corresponding pulse data for the Hadamard gate with a fidelity of 99.933\%. (c) Optimized pulse for the T-gate with 99.933\% gate fidelity. The Fourier transform shows a single peak centered at the $\ket{1}\leftrightarrow\ket{3}$ transition, serving to induce the required phase shift of $\pi/4$ for state $\ket{1}$.   
\label{fig:fluxonium_gates}}
\end{figure*}

Using closed-system optimal control, we optimize the control pulse $v(t)$ to realize three different single-qubit gates: the Pauli-X gate,  Hadamard gate, and the T gate,
\begin{equation*}
\mathrm{X} = \begin{pmatrix} 0 & 1 \\ 1 & 0 \end{pmatrix},\quad
\mathrm{H} = \frac{1}{\sqrt{2}}\begin{pmatrix} 1 & 1 \\ 1 & -1 \end{pmatrix},\quad
\mathrm{T} = \begin{pmatrix} 1 & 0 \\ 0 & e^{i\pi/4} \end{pmatrix}.
\end{equation*}
The latter two gates are known to form a universal set of single-qubit gates \cite{Nielsen2000}. Optimization must balance two conflicting requirements: gate times $t_g$ should be as short as possible to minimize the influence of dissipation and dephasing; at the same time, the maximum pulse amplitude $\max |v(t)|$ must remain small enough to avoid population of the resonator with unwanted photons. We find that pulses with $t_g$ on the order of a few tens of nanoseconds satisfy these conditions while also producing gates with high fidelities. In addition to the cost-functional contribution $C_1$, quantifying the target-gate infidelity, we employ additional cost contributions $C_3$ and $C_4$ to limit the time derivatives and maximum amplitude of the pulse $v(t)$. Suppressing the maximum amplitude ensures that occupation of the resonator with spurious photons is minimized. The cost on pulse derivatives helps eliminate unnecessary high-frequency components of $v(t)$ and render the pulses as smooth as possible, which is important for experimental applications, since instruments generating these control fields have a finite impulse response.

The pulses we obtain have a gate duration of $t_g = 60$\,ns  and closed system fidelities $>99.9\%$. The panels of
Fig.~\ref{fig:fluxonium_gates}(a)--(c) show the pulse $v(t)$ in the time domain and its discrete Fourier transform $\bar{v}(f)$ in the frequency domain. While interpreting optimized pulses is notoriously difficult, we note that general features of the three pulses and their frequency components can be given physical meaning. The Pauli-X and Hadamard gates both exhibit relatively well defined peaks in their Fourier spectra $\bar{v}(f)$ which coincide with the relevant transition frequencies among the lowest four levels primarily involved in the performance of the gate operation, see inset in Fig.\ \ref{fig:fluxonium_gates}(a). Visual inspection of $v(t)$ further reveals the staggered application of different frequency components. The initial and final $\sim5\,$ns time windows are dominated by high-frequency components related to transferring the system from the $\ket{0}$, $\ket{1}$ subspace to the delocalized states $\ket{2}$, $\ket{3}$ (and back). The central time window between $t=5$\,ns and $55$\,ns shows involvement of the low-frequency components associated with the transfer between the intermediary states $\ket{2}$ and $\ket{3}$. 
The T-gate, by contrast, exhibits a Fourier spectrum $\bar{v}(f)$ with only a single dominant frequency component corresponding to the  $\ket{1}\leftrightarrow\ket{3}$ transition. This is plausible, since the T-gate does not necessitate population transfer across the potential well. The transition peak for $\ket{1}\leftrightarrow\ket{3}$ facilitates the needed  $e^{i\pi/4}$ phase accumulation for the   $\ket{1}$ state.  

\begin{figure*}
  \includegraphics[width=\linewidth]{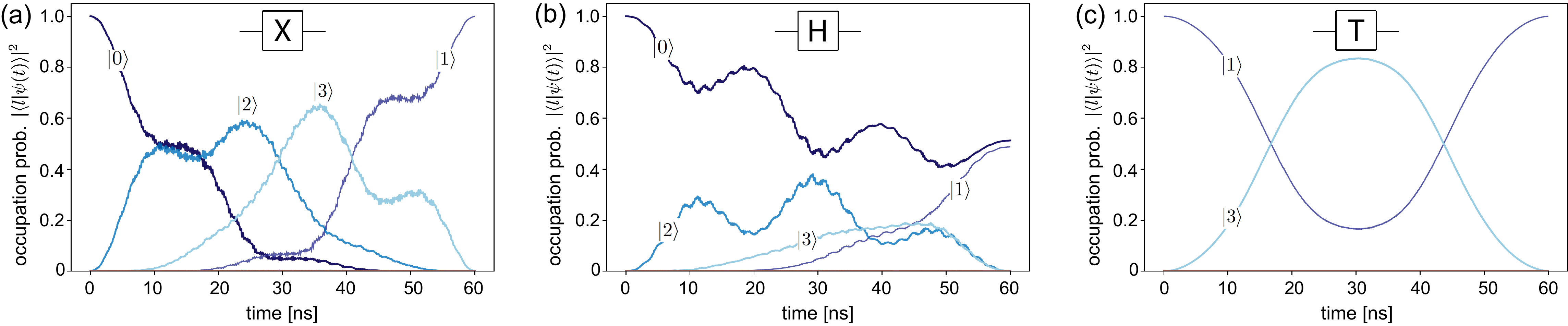}\\[-4mm]
  \caption{Time-evolution of state populations for 60\,ns high-fidelity single-qubit gates. (a) The time-evolution of states involved in the X gate shows state transfer between the qubit computational states via the delocalized $\ket{2}$ and $\ket{3}$ states. (b) For the H gate, states are transferred into an (approximately) equal superposition of $\ket{0}$ and $\ket{1}$. (c) In the T gate, the $\ket{1}$ state acquires an additional phase due to temporary state transfer into the $\ket{3}$ state.    
\label{fig:fluxonium_populations}}
\end{figure*}

Further evidence for this interpretation is given by Fig.\ \ref{fig:fluxonium_populations}, showing the probabilities for occupying the various fluxonium eigenstates as a function of time. For the Pauli-X gate and the Hadamard gate, occupation probabilities $p_l(t)=|\langle l | \psi(t)\rangle|^2$ are obtained for the example of initial qubit state $\ket{\psi(0)}=\ket{0}$. As expected, the X-gate transfers population into the final state $\ket{1}$, while the H-gate takes $\ket{0}$ into an equal superposition of $\ket{0}$ and $\ket{1}$. Both of these gates rely on the auxiliary states $\ket{2}$ and $\ket{3}$ to transfer population between the qubit compuational states. By contrast, the T-gate exhibits qualitatively dissimilar behavior since there is no need for state transfer across the fluxonium potential barrier. Instead, the much weaker control field only causes a small amount of intermediate population transfer from $\ket{1}$ to $\ket{3}$ for phase accumulation. 

Operating the fluxonium qubit away from its half-integer flux sweet spot makes the gate fidelity more vulnerable to the detrimental effects of $1/f$ flux noise. At an external flux of $0.45\Phi_0$, we expect flux noise to limit the dephasing time $T_\varphi$  and affect gate fidelities. To assess this issue, we follow a hybrid approach in which we evaluate dephasing rates due to classical $1/f$ noise, and then incorporate these rates in the Lindblad master equation. It must be emphasized that the latter step is a compromise we accept to avoid the heavier framework of non-Markovian master equations strictly appropriate for the inclusion of $1/f$ noise. This compromise is justified for a bound on the fidelity loss $\delta F$ in the present context, as gate durations are small compared to relevant dephasing times, $t_g\ll T_\varphi$. (Note that the exponential decay modeled by the Lindblad treatment is more rapid than the actual Gaussian decay at short times.)
Following Refs.~\cite{Ithier2005,Groszkowski_2018}, we consider the Gaussian decay (up to logarithmic corrections)  of the off-diagonal elements of the density matrix, and assign the standard deviation as the effective dephasing time. The leading-order result from inspection of the density-matrix element $\rho_{ll'}$ is given by
\begin{equation}\label{dephasing_rate}
    1/ (T^{\Phi_{\text{ext}}}_{\varphi})_{ll'} = \!A_{\Phi_{\text{ext}}} |\partial_{\Phi_{\text{ext}}}\omega_{ll'}| \sqrt{2|\text{ln}\,\omega_{\text{ir}}t|} .
\end{equation}
Here, $A_{\Phi_{\text{ext}}} = 1 \mu\Phi_0$ is the flux-noise amplitude \cite{Hutchings2017}, and $\omega_{ll'}$  the frequency difference between fluxonium states $\ket{l}$ and $\ket{l'}$. In our calculations, we use  $\omega_{\text{ir}}/2\pi = 1$\,Hz as the low-frequency cutoff,  and $t = 10\,\mu$s as the measurement time-scale \cite{Groszkowski_2018}.  For the heavy-fluxonium parameters stated above, the extracted dephasing rates are of the order of $\sim 1\, \mu$s, and  specifically $(T^{\Phi_{\text{ext}}}_{\varphi})_{10} = 3.1\, \mu$s for the two computational states.

Gate fidelities are also negatively affected by depolarization processes. Here, we consider dielectric surface loss as a likely candidate for limiting the $T_1$ time. While direct transitions among the computational states $\ket{0}$ and $\ket{1}$ are exponentially suppressed due to their disjoint support, transitions involving the delocalized levels $\ket{2}$ and $\ket{3}$ can occur. The corresponding transition rates are given by
\begin{equation}
    \gamma_{ll'} = \Gamma\, |\langle l | n_\phi |l'\rangle|^2,
\end{equation}
where we fix the rate constant $\Gamma$ by using the estimate $1/\gamma_{02} = 50 \mu$s, a realistic intra-well decay time observed in experiments using similar device parameters \cite{Premkumar2019}, and further supported by dielectric loss theory \cite{Nguyen2018}. 

Since the most relevant noise channels give rise to decoherence times that are about $10^2$ times larger than $t_g$, we expect that open-system simulation using the optimized pulses should only lead to small changes in gate fidelities. In our calculation, we use a master equation of the form
\begin{equation}\label{master_equation}
    \frac{d\rho(t)}{dt} = -i[H_f + V(t), \rho(t)] + \big(\mathbb{D}[c_0] + \sum_{l<l'}\mathbb{D}[c_{ll'}]\big)\rho(t),
\end{equation}
where dephasing due to flux noise is captured by the diagonal jump operator $c_0 = \sum_{l}\sqrt{(\gamma_\varphi)_{l0}}\ket{l}\!\bra{l}$, and depolarization due to dielectric loss by $c_{ll'} = \sqrt{\gamma_{ll'}}\ket{l}\!\bra{l'}$. The Lindblad damping superoperator has the usual form $\mathbb{D}[c]\rho = c\rho c^\dag - \frac{1}{2}\{c^\dag c,\rho\}$. We calculate the resulting open-system gate fidelities by means of the expressions detailed in Appendix \ref{fidelitycalculation}. The resulting open-system fidelities for the X, H, and T gates are 99.66\%, 99.60\%, and 99.59\%, respectively. This should be compared to the corresponding closed-system fidelities of 99.94\%, 99.93\%, and 99.93\%.

\subsection{Controlled-Z gate}
To obtain a set of gates universal for multi-qubit unitaries, we demonstrate an optimized controlled-Z (CZ) gate,
\begin{equation*}
\mathrm{CZ} = \begin{pmatrix} 1 & 0 & 0 & 0 \\ 0 & 1 & 0 & 0 \\ 0 & 0 & 1 & 0 \\ 0 & 0 & 0 & -1 \end{pmatrix},
\end{equation*}
with a gate time of $t_g = 60$\,ns. We consider two heavy-fluxonium qubits with identical circuit parameters of the same values as in section \ref{single_qubit_gates}, but biased by different magnetic fluxes $\Phi_{\text{ext},1} = 0.45\Phi_0$ (target qubit) and $\Phi_{\text{ext},2} = 0.455\Phi_0$ (control qubit). Biasing the qubits by this small flux offset results in coupling-induced energy shifts of $\sim$ 10 MHz for higher level, non-computational states, which can be used for state entanglement. The interaction leaves computational states essentially unshifted due to suppression of the $\langle 0 | n_\phi| 1\rangle$ matrix element. 

We couple each qubit to a shared resonator through a small coupling capacitor. The resulting Hamiltonian 
\begin{equation}
    H(t) = H^{(1)}_{f} + H^{(2)}_{f} + H^{(1,2)} + V^{(1)}(t) + V^{(2)}(t)  
\end{equation}
generalizes Eq.\ $\eqref{total_Ham_single_qubit}$. Here, $H^{(i)}_{f}$ are the Hamiltonians for the two fluxonia ($i=1,2$), and $V^{(i)}(t)$ is the dispersively filtered drive acting on each, in the form given in Eq.\ $\eqref{filtered_drive}$. The qubits are driven by two separate pulses $v(t)$ and $w(t)$ for target and control respectively. Due to the coupling to a shared resonator, there is an effective mutual coupling $H^{(1,2)}$ between the two fluxonium qubits that allows for entanglement generation, see Appendix \ref{dispersive} for details.

\begin{figure*}
  \includegraphics[width=\linewidth]{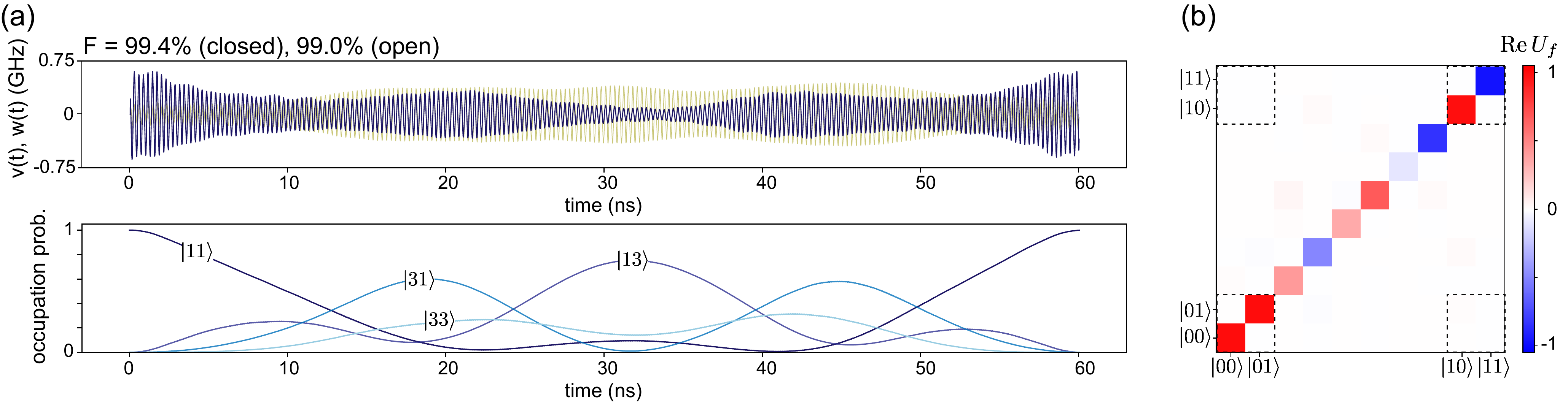}\\[-4mm]
  \caption{Controlled-Z gate for two heavy-fluxonium qubits with a gate time of 60\,ns [$\Phi_{\text{ext},1} = 0.45\Phi_0$ (target qubit) and $\Phi_{\text{ext},2} = 0.455\Phi_0$ (control qubit)]. (a) The top panel shows optimized pulses acting on the target and control qubit, $v(t)$ and $w(t)$, respectively, achieving a closed-system fidelity of 99.4\% and open-system fidelity of 99.0\%. The cost functionals used are $C_1$, $C_3$, and $C_4$. The bottom panel shows occupation probabilities of system eigenstates $\ket{ml}$, with $\ket{11}$ chosen as the initial state. $\ket{11}$ undergoes the significant population to intermediate states so to induce a phase of $e^{i\pi}$, as required for the CZ gate.  (b) Real part of the resulting unitary $U_f$ achieved by optimization, showing levels $\ket{0l}$ with $0\leq l \leq 7$, $\ket{10}$, and $\ket{11}$. Matrix elements between states in the computational subspace are marked by dashed squares. 
\label{fig:fluxonium_two_qubits}}
\end{figure*}

As shown in Fig.\ \ref{fig:fluxonium_two_qubits}(a),  optimal-control theory yields pulses $v(t)$ and $w(t)$ that activate a two-fluxonium CZ gate. As in the case of a single fluxonium qubit, we employ cost contributions $C_1$, $C_3$, and $C_4$ in the optimization. The bottom panel of Fig.\ \ref{fig:fluxonium_two_qubits}(a) monitors the system time evolution in terms of the occupation probabilities of participating states. We have confirmed that the case with $\ket{11}$ acting as initial state shows the largest amount of intermediate population transfer. This is consistent with the fact that this state must acquire a phase factor of $e^{i\pi}$, accomplished by the observed excursion into states $\ket{13}$, $\ket{31}$, and $\ket{33}$. Like the T gate, only transitions between states $\ket{1}$ and $\ket{3}$ are necessary to accumulate phase factors. Fig.\ \ref{fig:fluxonium_two_qubits}(b) depicts the gate unitary achieved by optimization. $U_f$ is represented in the product basis $\ket{ml}$, with $m$ and $l$ labeling control and target qubit levels, respectively. The relevant elements in the 4$\times$4 computational subspace are marked by dashed rectangles, and have entries which closely match the controlled-Z target unitary. Overall, the optimized pulse trains realize the CZ gate with a closed-system fidelity of 99.4\% using the same gate time $t_g = 60$\,ns. Open-system simulations including noise contributions from 1/$f$ flux noise and dielectric loss result in an open-system fidelity of 99.0\%.

Our optimal-control results for the CZ gate may be compared to the recent work by Nesterov et al.\ \cite{Nesterov2018}. The setup in that work differs in the utilization of \emph{direct} capacitive or inductive coupling between the fluxonia, which are then driven without shared resonator by a microwave tone with a Gaussian envelope. The pulse is optimized over the amplitude and drive frequency, rather than using a general-purpose optimal-control package. For the same gate time of 60\,ns and direct capacitive coupling, they report a similar closed-system fidelity of 99.3\% employing a single pulse on only one qubit.

\section{Optimized single-qubit gate set for the \zeropi qubit}
The \zeropi circuit \cite{Brooks2013,Dempster2014,Groszkowski_2018,DiPaolo2019} further extends the protection afforded by the heavy-fluxonium qubit by combining exponential suppression of \emph{both} relaxation and dephasing due to disjoint wave-function support and robust ground-state degeneracy. 
We briefly review the physics of the \zeropi circuit and the parameters required for its protected regime. The circuit consists of two superinductors of inductance $E_L$, two Josephson junctions with Josephson energy $E_J$ and junction capacitance $C_J$, and two large shunt capacitors $C$. The ideal \zeropi Hamiltonian reads
\begin{align}
    H_{0-\pi} &= \frac{(q_{\theta}-n_g)^2}{2C_\theta} + \frac{q_\phi^2}{2C_\phi}  \\\nonumber
    &\quad + E_L\phi^2 - 2E_J\,\text{cos}(\theta)\text{cos}(\phi - \pi\Phi_{\text{ext}}/\Phi_0),
\end{align}
where $n_g$ is the offset charge, and  $q_\theta = 2en_\theta$, $q_\phi = 2en_\phi$ are the charge operators canonically conjugate to the two degrees of freedom $\theta$ and $\phi$. 
The effective capacitances associated with these two variables are $C_\theta = 2(C + C_J) + C_g$ and $C_\phi = 2C_J + C_g$, where $C_g$ is a small capacitance due to coupling to ground and external voltage lines \cite{DiPaolo2019}. The external magnetic flux threading the circuit loop is denoted $\Phi_{\text{ext}}$.

\begin{figure*}
  \includegraphics[width=0.9\textwidth]{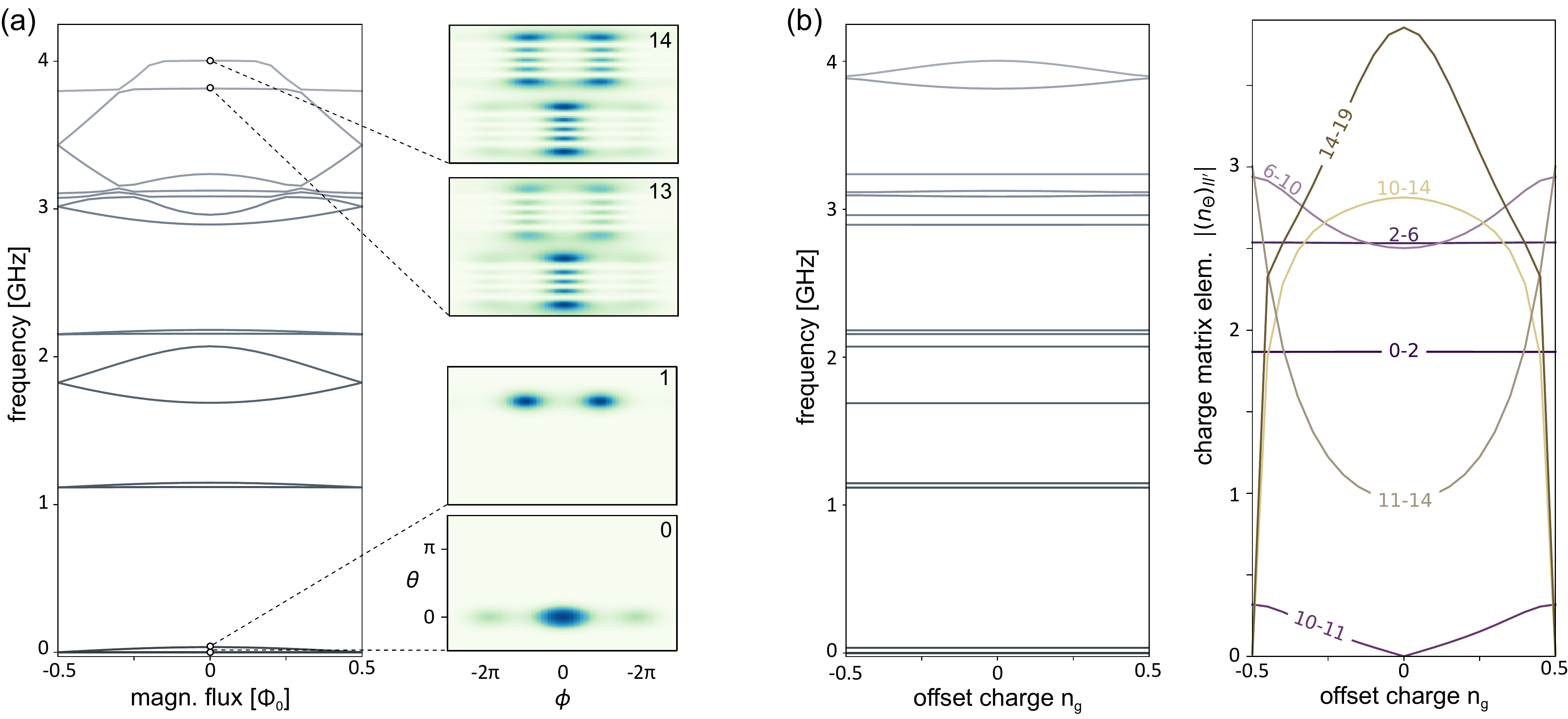}\\[-2mm]
  \caption{Spectrum and matrix elements of the \zeropi qubit. (a) Lowest 14 eigenenergies vs.\ magnetic flux,  and select \zeropi eigenfunctions at $\Phi_{\text{ext}} = 0$ and offset charge $n_g = 0.25$. The lowest two eigenfunctions, forming the computational subspace, are localized around $\theta=0$ and $\theta=\pi$, and are nearly degenerate. The top two eigenfunctions shown are delocalized states which occupy both potential wells, and serve as auxiliary states in gate operations.
  (b) Eigenenergies and $n_\theta$ charge matrix elements vs.\ offset charge $n_g$. As eigenstates start delocalizing along $\theta$, eigenenergies become weakly dependent on offset charge. Charge matrix elements show significant dependence on $n_g$  for transitions between delocalized high-energy states, e.g. $\ket{11} \rightarrow \ket{14}$. 
    (Parameters: $E_L/h = E_C/h = 40$\,MHz, $E_J/h = 10$\,GHz, and $E_{CJ}/h = 20$\,GHz.)
\label{fig:flux_and_ng}}
\end{figure*}

For the \zeropi qubit to realize the desired intrinsic protection, circuit parameters must satisfy several conditions. To achieve localization along the $\theta$ axis, the effective mass in $\theta$ needs to be heavy compared to that in $\phi$ direction, $C_\theta \ll C_\phi$, and the local potential wells deep enough to hold localized states, $e^2/2C_\theta\ll E_J$. The latter condition also renders the qubit charge-noise insensitive. To suppress sensitivity to flux noise, wave functions should be delocalized along the $\phi$ axis, obtained when $E_L\ll E_J$,\, $e^2/2C_\phi$. Achieving this parameter regime remains experimentally challenging. Here, we choose an ``optimistic" parameter set previously considered in  Groszkowski et al.\ \cite{Groszkowski_2018}, namely $E_L/h = E_C/h = 40$\,MHz, $E_J/h = 10$\,GHz, and $E_{CJ}/h = 20$\,GHz.  This choice provides an appropriate amount of qubit protection. The eigenspectrum as a function of external flux $\Phi_{\text{ext}}$ is shown in Fig.\ \ref{fig:flux_and_ng}(a) along with several eigenfunctions. The lowest two, $\ket{0}$ and $\ket{1}$, span the computational subspace and are localized along $\theta = 0$ and $\theta = \pi$, respectively. States higher up in the spectrum, such as $\ket{13}$ and $\ket{14}$, are delocalized in the $\theta$ direction and will play an important role in the gate protocols.

Similar to the situation with the heavy-fluxonium qubit, disjoint support of the computational basis states in the \zeropi qubit provides intrinsic protection from decoherence, but inevitably also prevents one from driving direct transitions between the two qubit states. Di Paolo et al.\ \cite{DiPaolo2019} achieved gate operations between states indirectly via a square voltage pulse that drives transitions via intermediate higher excited levels. Depending on device parameters, this strategy results either in an X gate or a Hadamard gate, but does not readily yield a gate set universal for single-qubit operations. For our optimal-control search, we consider the more conventional method of dispersively coupling the \zeropi qubit to a resonator via $n_\theta$, and driving the qubit via this resonator with a microwave pulse. 
Together, the drift and control Hamiltonian for \zeropi acquire a form analogous to that encountered for heavy fluxonium in the previous section,
\begin{equation}
    H(t) = H_{0\text{-}\pi}(n_g) + V(t).
\end{equation}
The control Hamiltonian $V(t)$ takes the form of Eq.\ $\eqref{filtered_drive}$, in which we take the filtered drive to couple to the $\theta$ degree of freedom (i.e., $n_\phi$ is replaced with $n_\theta$). 
Employing optimal control to find the appropriate pulse shapes $v(t)$ gives sufficient flexibility for realizing a variety of single-qubit gates.

However, one challenge concerning \zeropi gates which has not previously been discussed is revealed by Fig.\ \ref{fig:flux_and_ng}(b), showing the dependence of charge matrix elements $(n_\theta)_{jj'}=\langle j | n_\theta | j'\rangle$ on the offset charge $n_g$. Among low-lying, $\theta$-localized states, these matrix elements are practically $n_g$-insensitive as expected. By contrast, as higher-energy states start delocalizing in the $\theta$ direction, offset-charge dependence of matrix elements becomes significant. This offset-charge sensitivity may affect gate operations which utilize higher-energy states as a means to transfer probability amplitude between the $\theta=0$ and $\theta=\pi$ wells. The problem is exacerbated by the fact that offset charge is not controlled in experiments and is subject to significant fluctuations due to 1/$f$ charge noise \cite{Zorin1996,Pourkabirian2014}. Our strategy is thus to steer the optimizer towards control solutions that are maximally insensitive to offset-charge fluctuations. We have enhanced the optimal-control code to allow for drift and control Hamiltonians to vary from iteration to iteration, allowing us to choose random values of offset charge (using a uniform distribution over $0\le n_g < 1$) for each individual iteration of the optimizer. 
Directly applying the gradients from each iteration results in a stochastic-gradient-descent  \cite{Ketkar_2017,Zinkevich_Weimer_Li_Smola_2010} process. With careful tuning of cost-function weights, this process converges to an average solution balancing all possible values of $n_g$.
A second challenge concerns the inevitable presence of disorder in circuit components which can lead to spurious coupling to a harmonic, low-energy degree of freedom, the $\zeta$-mode \cite{Dempster2014,Groszkowski_2018,DiPaolo2019}.
To avoid the overhead of a significant increase in Hilbert space dimension, we apply the optimal-control formalism to the ideal \zeropi system, and verify subsequently that weak coupling to the $\zeta$-mode does not significantly reduce gate fidelities.

 \begin{figure*}
  \includegraphics[width=0.9\linewidth]{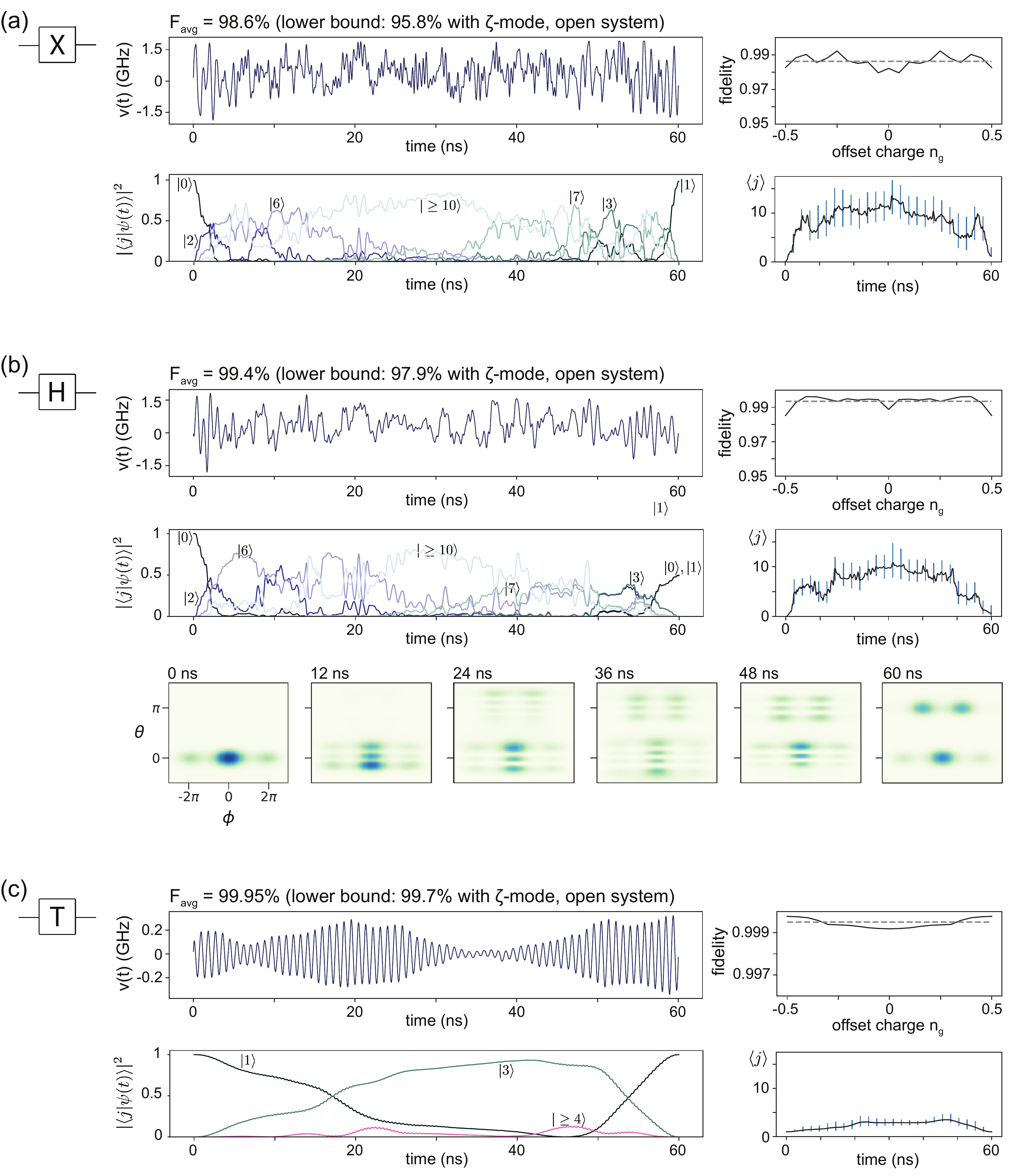}\\[-4mm]
  \caption{Optimized pulses for \zeropi single-qubit gates at $\Phi_{\text{ext}} = 0$. For X gate, Hadamard gate and T gate, (a)-(c) show panels with pulse trains $v(t)$, fidelities vs.\ offset charge, occupation probabilities, and mean and standard deviation of occupied levels (using $n_g=0.25$). Fidelity variations vs. offset charge are observed to be small compared to the average fidelity. The additional bottom panel in (b) depicts time evolution snapshots of \zeropi wave functions for the optimized Hadamard gate.
\label{fig:zeropi_gates}}
\end{figure*}

Figure \ref{fig:zeropi_gates} presents the results from optimal-control theory for the three single-qubit gates  X, H, and T. For all \zeropi gates, we again choose a gate duration of $t_g = 60$\,ns which aims to balance, on one hand, gate fidelity benefiting from short gate times; and on the other hand, overall pulse power which decreases as gate duration is increased. Both the X and Hadamard gate [Fig.\ \ref{fig:zeropi_gates}(a,b)] require probability-amplitude transfer between computational-basis states, and are seen to result in similar level-population dynamics accessing excited states $|{\ge}10\rangle$. Delocalization of these states in the $\theta$ variable enables the population transfer between the computational states, but also temporarily lifts the protection granted by disjoint-support wavefunctions. 

For both X and H gates, the plots of occupation probability amplitude $p_j=|\langle j |\psi(t)\rangle|^2$ show initial and final phases of state transfer out of and into the computational-basis subspace. The sequences of either even or odd-numbered levels reflect transitions among states centered at $\theta=0$ and $\pi$, respectively. (The states $\ket{4}$ and $\ket{5}$ do not contribute due to the lack of connecting matrix elements.) Beyond these initial and final phases, the dynamics is dominated by an extended intermediate phase during which higher-excited states participate. As expected, these states are delocalized in $\theta$ as illustrated by the intermediate-time evolution snapshots for the Hadamard gate [Fig.\ \ref{fig:zeropi_gates}(b)].  As depicted in the plots of the average level occupied and the standard deviation (calculated separately for levels above and below the average), the intermediate-phase dynamics is not readily interpreted as a sequence of transitions among higher levels, but rather involves evolution among superpositions of such states.   

Pulse shape and population dynamics are qualitatively different for the T gate which does not require transfer of amplitude from the $\theta=0$ to the $\theta=\pi$ well, see Fig.\ \ref{fig:zeropi_gates}(c). Instead, the $e^{i\pi/4}$ phase accumulation required for the $\ket{1}$ state is obtained by temporary occupation of state $\ket{3}$ and subsequent back-transfer into $\ket{1}$.

The closed-system fidelities, averaged over offset charge $n_g$, are 98.6\% for the X gate, 99.4\% for the H gate, and 99.95\% for the T gate. The sub-99.9\% fidelities for the X and H gates can be attributed to the significant offset-charge dependence of the transition matrix elements $(n_\theta)_{jj'}$ between high-energy delocalized states, see Fig.\ \ref{fig:flux_and_ng}(b). Since pulses for X and H gates require  occupation of these high-energy auxiliary states intermediately, optimization must compromise between higher-fidelity solutions for different fixed offset charges $n_g$. The trade-off is most significant between the offset charge values $n_g = 0$ and $0.5$.  As shown in Fig.\ \ref{fig:flux_and_ng}(b), certain transitions between delocalized states are symmetry-forbidden at $n_g = 0$, such as the $\ket{10}\rightarrow\ket{11}$ transition. This forces the system to take a detour into higher-energy states such as $\ket{13}$ or $\ket{14}$. The situation is reversed for $n_g = 0.5$, where $\ket{10}\rightarrow\ket{11}$ is allowed, but $\ket{10}\rightarrow\ket{14}$ is strongly suppressed. The fidelity plots for the X and H gates in Fig.\ \ref{fig:zeropi_gates}(a,b) are consistent with this trade-off, showing decreases in fidelity at $n_g=0$ and $0.5$. The resulting pulse shapes appear more complicated than their heavy-fluxonium counterparts and their Fourier transforms (not shown) do not exhibit the well-resolved peaks observed in the fluxonium case.

We next investigate the performance of the optimized pulses in the presence of circuit disorder -- in particular, in $C$ and $L$. Such disorder leads to spurious coupling to the  harmonic $\zeta$-mode with mode frequency $\Omega_\zeta = \sqrt{8E_LE_{C_\zeta}}/\hbar$, where $E_{C_\zeta} = e^2/2C_\zeta$ and $C_\zeta = 2C + C_g\approx 2C$ \cite{DiPaolo2019,Groszkowski_2018}. The resulting Hamiltonian reads
\begin{equation}
    H = H_{0-\pi} + \Omega_\zeta a^\dag\!a + \sum_{j,j'}(\kappa_{jj'}\ket{j}\!\bra{j'}a + \text{h.c.}),  
\end{equation}
where $\ket{j}$ are \zeropi eigenstates and $\kappa_{jj'} = \kappa^\phi_{jj'} + i\,\kappa^\theta_{jj'}$ are coupling strengths defined by
\begin{align}
\kappa^\theta_{jj'} &= \frac{1}{2}E_{C\Sigma}\,dC \left(\frac{32E_L}{E_{C_\zeta}}\right)^{1/4}\bra{j}\!n_\theta\!\ket{j'},\\
\kappa^\phi_{jj'} &= \frac{1}{2}E_LdE_L\left(\frac{8E_{C_\zeta}}{E_L}\right)^{1/4}\bra{j}\!\phi\!\ket{j'}.
\end{align}
We assume the relative disorder in inductance and capacitance to be at the level of $dL = dC = 5\%$. Disorder also leads to an additional component to the drive that couples to the $\zeta$-mode,
\begin{equation}
  n_\theta \rightarrow n_\theta - \beta n_\zeta,  
\end{equation}
where $n_\zeta$ is the charge-number operator for the $\zeta$-mode and $\beta = C\,dC/C_\zeta$ (see App.\ A of Ref.\ \onlinecite{DiPaolo2019}).

As noted in Ref.\ \onlinecite{Groszkowski_2018}, the coupling to the $\zeta$-mode opens up an unwanted shot-noise dephasing channel that is absent in the symmetric \zeropi device, and can become a dominant source of dephasing. 
In addition to shot-noise, we consider 1/$f$ charge noise. Since the induced dephasing rates are offset-charge dependent, we consider a worst-case scenario by maximizing dephasing rates over $n_g$. Charge noise impacts the system at intermediate times of the gate protocol, when the system occupies unprotected high-energy states that are delocalized in the $\theta$ variable. We further take into account the effect of dissipation  due to  dielectric surface loss. Like in heavy-fluxonium, while direct transitions between $\ket{1}$ and $\ket{0}$ are suppressed due to disjoint support, transitions between excited states can still occur. Using realistic parameters and operating at the $\Phi_{\text{ext}} = 0$ sweet spot, we have confirmed that other noise channels such as critical-current fluctuations and flux noise are subdominant and do not lead to significant reductions of the fidelity. 

For the open-system simulation, we employ the obtained pulses and evolve the system composed of \zeropi and $\zeta$-mode under the Lindblad master equation [see Eq.\, \eqref{master_equation}]. Shot noise is incorporated using absorption and relaxation rates $\kappa_\zeta n_{\text{th}}(\Omega_\zeta)$ and $\kappa_\zeta n_{\text{th}}(\Omega_\zeta) + 1$ respectively, where the $\zeta$-mode thermal occupation $n_{\text{th}}(\Omega_\zeta)$ is 2.29 at a temperature of 15\,mK and $1/\kappa_\zeta = 100$\,$\mu$s. Charge-noise dephasing is incorporated using dephasing rates $(\gamma_\varphi)_{j0}$, evaluated using the analogue of Eq.\, \eqref{dephasing_rate}, here with derivative evaluated with respect to $n_g$. 
Finally, we model dielectric decay rates in the same way as for fluxonium, i.e., we take $\gamma_{jj'} = \Gamma\, |\langle j | n_\theta |j'\rangle|^2$ and fix the rate constant $\Gamma$ by taking $1/\gamma_{02} = 50\mu$s. 

Based on this master-equation simulation, we obtain conservative lower bounds on average gate fidelities for the three single-qubit gates: 95.8\% for X, 97.9\% for H, and 99.7\% for T. Evidently, inclusion of noise and coupling to the $\zeta$-mode mainly affects the X and H gates. This fidelity loss is primarily due to two factors: shot-noise dephasing induced by the $\zeta$-mode, as well as occupation of higher delocalized states with enhanced sensitivity to charge noise. (Note that lower bounds report the worst fidelity, reached for a particular $n_g$.)  We thus find that the main barrier to achieving high-fidelity single-qubit gates for \zeropi is coupling to the $\zeta$-mode, offset-charge fluctuations, and dielectric surface loss.

\section{Conclusion}

In summary, we have used optimal control theory for implementing quantum gates for protected superconducting qubits whose computational states have practically disjoint support. We have presented optimal control pulses yielding a fully universal set of gates for the heavy-fluxonium qubit and a set universal for single-qubit gates for the \zeropi qubit. Specifically, we considered the Pauli-X gate, Hadamard gate, and T gate with closed-system fidelities of $>$99.9\% for heavy-fluxonium, likewise a controlled-Z gate with a closed-system fidelity of 99.4\%. For the \zeropi qubit, we implemented an enhanced optimal-control method by allowing offset charge to vary for each optimizer iteration. Applying gradients from each iteration results in a stochastic-gradient-descent process. This process converges to an average solution yielding a fidelity averaged over a range of $n_g$. We presented pulses with closed-system average fidelities of 98.6\%, 99.4\%, and 99.95\% for X, H, and T respectively. Remarkably, this method thus provides a way to find control pulses with good fidelities which are roughly insensitive to random offset charge changes.  

All constructed gates represent a compromise between limiting drive powers to realistic values and mitigating the effects of noise. To assess the fidelity losses in the open system, we incorporated  optimized pulses into a master equation treatment. For heavy-fluxonium, 1/$f$ flux dephasing and dielectric surface losses were the primary sources of decoherence. The resulting open-system fidelities obtained were $>$99\% for single-qubit gates and 99.0\% for the controlled-Z gate. For \zeropi, shot-noise dephasing in the $\zeta$-mode, 1$/f$ offset-charge dephasing, and dielectric surface loss were the most relevant noise sources. The resulting conservative lower bounds on average fidelities for \zeropi + $\zeta$-mode were 95.8\%, 97.9\%, and 99.7\% for X, H, and T. 

Future work should consider the possibility of combining drive pulses with the active cooling scheme proposed in Ref.\ \cite{DiPaolo2019}, where it was discussed primarily for the qubit idling. Further extensions may also entail using open-system optimization algorithms \cite{Open-Mohamed,boutin2017resonator} to partially mitigate gate-fidelity losses due to noise. This will involve significantly more computational overhead. We believe that optimal control provides a promising avenue toward universal gates on today's protected superconducting qubits.

\begin{acknowledgements}
We thank Peter Groszkowski, Anjali Premkumar, and Agustin Di Paolo for valuable discussions. We gratefully acknowledge support from the Army Research Office through Grant Nos.\ W911NF-15-1-0421 and W911NF-19-1-0016.
\end{acknowledgements}

\vspace*{3mm}
\noindent
M.A.\ and B.B.\ contributed equally to this work.

\appendix

\section{Open-System Gate Fidelity\label{fidelitycalculation}}
To assess the effects of dissipation on optimized gates, we employ a definition of open-system fidelity that correctly reduces to the open-system fidelity when the coupling to the environment is eliminated. This allows for consistent comparison open and closed-system fidelities. The definition makes use of the density matrix in vectorized (coherence vector) form, stacking the rows of the $n{\times}n$ density matrix $\rho$ into a $n^2{\times}1$  vector $\dket{\rho}$,
\begin{equation}
\begin{split}
    \rho = \sum_{ij} \rho_{ij} \ket{i}\bra{j} \longrightarrow &\dket{\rho} =  \sum_{ij} \rho_{ij} \ket{i} \ket{j}. 
\end{split}
\end{equation}
The evolution of the density matrix is then written with help of the $n^2{\times}n^2$ superoperator $L$ as $\dket{\rho(t)} = L \dket{\rho(0)}$.

A consistent measure for the open-system gate fidelity is then given by
\begin{equation}
\label{eq:def_fid}
    F_o = \frac{1}{n^2} \text{Tr} (L_t^{\dagger} L_f),
\end{equation}
where $L_t$ is the target superoperator and $L_f$ is the final achieved superoperator.
This metric correctly reduces to the expression of the closed-system trace fidelity $F_c = |\text{Tr}(U_t^{\dagger} U_f/n)|^2$ when dissipation is switched off. To prove this, we first describe the action on vectorized states equivalent to matrix multiplication from the right and left on $\rho$,
\begin{align*}
    A \rho &= \sum_{ij} \rho_{ij} A\ket{i}\bra{j} \longrightarrow \sum_{ij}\rho_{ij} A\ket{i} \ket{j} = A \otimes I \dket{\rho},\\
     \rho B &= \sum_{ij}\rho_{ij} \ket{i} \bra{j} B \longrightarrow \sum_{ij} \rho_{ij} \ket{i} B^T\ket{j} = I \otimes B^T \dket{\rho}.
\end{align*}
In the case of closed-system dynamics, the density matrix is a pure state $|\psi\rangle\langle\psi|$, and the evolution reduces to 
\begin{equation}
    L_c \rho = L_c \ket{\psi}\bra{\psi} = U\ket{\psi}\bra{\psi}U^{\dagger} = U \rho U^{\dagger},
\end{equation}
where $U$ is the closed-system propagator. In the vectorized picture, this reads
\begin{equation}
L_c \rho = U \rho U^{\dagger} \longrightarrow{}  U \otimes U^{*} \dket{\rho}.
\end{equation}
Therefore, computation of $F_o$ yields
\begin{equation}
    F_o = \frac{1}{n^2} \text{Tr}(U_t^{\dagger} U_f \otimes U_t^{T} U_f^{*}) = \frac{1}{n^2} |\text{Tr}(U_t^{\dagger} U_f)|^2 = F_c,
\end{equation}
consistent with the closed-system gate fidelity.

\section{Dispersively Filtered Drive and Qudit Coupling\label{dispersive}}

This appendix briefly summarizes the derivation of the dispersive drive term [Eq.\ \eqref{filtered_drive}] and the effective qudit-qudit coupling mediated by a resonator. The derivation follows as a slight generalization from Ref.\ \cite{Zhu2013}, and is based on the dispersive Schrieffer-Wolff transformation.

Consider a generalized Jaynes-Cummings Hamiltonian describing two qudits coupled to a resonator, the latter  driven by a microwave tone $v(t)$:
\begin{equation}
     H = H_0 + V + v(t)(a^\dag + a).    
\end{equation}
Here, $H_0=\omega_r a^\dag a + H_0^{(1)} + H_0^{(2)}$ comprises the resonator and bare qudit Hamiltonians. $V=V^{(1)} + V^{(2)}$ describes the qudit-resonator coupling which is of the form
\begin{equation}\label{coupling}
V^{(k)} = \sum_{ll'}g^{(k)}_{ll'}\ket{l_k}\!\bra{l_k'}(a^\dag + a).
\end{equation}
$\ket{l_k}$ are the eigenstates of qudit $k$, and $g^{(k)}_{ll'} = g\bra{l_k}\!n_\phi\!\ket{l'_k}$ the coupling matrix elements with overall strength $g$. In the dispersive regime, detunings are large compared to the coupling, i.e.,  $\lambda_{ll'}^{(k)} = g_{ll'}^{(k)}/\Delta_{ll'}^{(k)} \ll 1$ with $\Delta_{ll'}^{(k)} = \epsilon_l^{(k)} - \epsilon_{l'}^{(k)} - \omega_r$ denoting the detuning between qudit-$k$ transition $l\to l'$ and the resonator, and $\epsilon_l^{(k)}$ the $l$-th eigenenergy of qudit $k$. For our second-order treatment, we only require the leading order of the Schrieffer-Wolff transformation generator,
\begin{equation}\label{first_order_generator}
    S \equiv -i\sum_k\sum_{ll'}(\lambda_{ll'}^{(k)}a - \lambda_{l'l}^{(k)}a^\dag)\ket{l_k}\!\bra{l'_k}.
\end{equation}

with the second-order result
\begin{align}\label{effective_JC_two_qudits}
    H' &= H_0 + v(t)(a^\dag + a) + V +  [iS, H_0 + v(t)(a^\dag + a)]\nonumber \\ 
    & \quad + [iS, V] +  \frac{1}{2}[S,[S,H_0 + v(t)(a^\dag + a)]].
\end{align}

First, focus on the first-order term $[iS^{(k)}, v(t)(a^\dag + a)]$. This term is the leading drive contribution on qudit $k$. Evaluating the commutator yields 
\begin{equation*}
[iS^{(k)}, v(t)(a^\dag + a)] = 2g\omega_r v(t) \sum_{ll'}\frac{\ket{l_k}\bra{l_k}n_\phi\ket{l'_k}\bra{l'_k}}{(\epsilon_{l}^{(k)} - \epsilon_{l'}^{(k)})^2-\omega_r^2},     
\end{equation*}
which captures the filtered drive on qudit $k$.

Second-order terms in Eq.\, \eqref{effective_JC_two_qudits} become essential for calculation of the effective qudit-qudit coupling. The terms that lead to coupling are the ``cross" commutators, $[iS^{(k)}, V^{(j)}]$ and $\frac{1}{2}[S^{(k)},[S^{(j)},H_0]]$ ($k\neq j$). Together, these generate the coupling term
\begin{equation}
    H^{(1,2)} = \frac{g^2}{2}\left( n_\phi^{(1)}\otimes \widetilde{n}_\phi^{(2)} + \widetilde{n}_\phi^{(1)}\otimes n_\phi^{(2)}\right),
\end{equation}
where
\begin{equation}
    \widetilde{n}_\phi^{(k)} \equiv 2\omega_r\sum_{ll'}\frac{\bra{l_k}n_\phi\ket{l'_k}}{(\epsilon_{l}^{(k)} - \epsilon_{l'}^{(k)})^2-\omega_r^2}\ket{l_k}\!\bra{l'_k}.
\end{equation}
The above coupling term is the natural generalization of the two-qubit coupling discussed in Ref.\ \onlinecite{Majer2007}.

\bibliographystyle{apsrev4-1}
\bibliography{paper_refs} 

\end{document}